# A Data-Driven Residential Transformer Overloading Risk Assessment Method

Ming Dong, *Senior Member, IEEE*, Alexandre B. Nassif, *Senior Member, IEEE*, and Benzhe Li, *Member, IEEE*

***Abstract*—Residential transformer population is a critical type of asset that many electric utility companies have been attempting to manage proactively and effectively to reduce unexpected transformer failures and life loss that are often caused by overloading. Within the typical power asset portfolio, the residential transformer asset is often large in population, has the lowest reliability design, lacks transformer loading data, and is susceptible to customer loading behaviors, such as adoption of distributed energy resources and electric vehicles. On the bright side, the availability of more residential service operation data along with the advancement of data analytics techniques has provided a new path to further our understanding of residential transformer overloading risk statistically. This paper develops a new data-driven method that combines transformer temperature rise and insulation life loss simulation model with clustering analysis technique. It quantitatively and statistically assesses the overloading risk of residential transformer population in one area and suggests proper risk management measures according to the assessment results. Multiple application examples for a Canadian utility company have been presented and discussed in detail to demonstrate the applicability and usefulness of the proposed method.**

## I. Introduction

NOWADAYS, many electric utility companies are seeking ways to understand and manage their power asset risk. Under deregulated environment, electric utilities are encouraged to reduce overall cost while maintaining system reliability risk at an acceptable level. Within the typical power asset portfolio, the effective assessment and management of residential transformer risk of failure has become particularly important due to the following reasons:

- Residential transformers are often large in population, especially in urban utilities. As a result, failures of residential transformers are practically difficult to avoid.
- Residential transformers are normally not designed for the N − 1 reliability requirement which upper stream of power delivery systems, such as distribution feeders, substation transformers and breakers are required to comply with [1].
- This means the failure of a residential transformer often results in hours of effort for the replacement to restore power supply to customers. This duration could sometimes lead to severe consequence for residential customers, especially in extreme weather conditions.
- The reliability risk of residential transformers is directly affected by the loading behaviors of residential customers [2]–[4]. In recent years, the adoption of distributed energy resources (DERs) and electric vehicles (EVs) further increased the uncertainty of the aggregated load behavior at the residential transformer level as the uptake trend continues and the customer adoption level has not entered into a steady stage [5]–[7].

Unlike the risk assessment of power transformers which has been researched extensively [8]–[12], it is not costly justifiable to install health monitoring devices on those prevalent and low-cost residential transformers and assess risk based on monitoring results. Even the load profiles of residential transformers are not easy to measure directly. The risk assessment of residential transformers therefore should aim at developing a statistical assessment and management approach for the entire population which many utility companies are not familiar with.

In spite of the above challenges, developments in other areas have also enabled new ways to residential transformer risk assessment and management. During the past decade, the deployment of smart meters increased the visibility of individual residential services' loading behavior. In places where smart meters have not been installed, interval meter data are often made available to support statistical analysis such as load settlement [13], [14]. These residential service loading data can be aggregated to the transformer level using a bottom-up approach. In addition to metering data, other residential operation related data such as historical weather and temperature data have become readily available in today's big data era [15], [16]. Meanwhile, advanced data analytics methods have been applied to power systems. For example, [17] uses a database to store typical distribution transformer load profiles and the corresponding calculated life loss. Then a new load profile can be classified with respect to the database profiles based on minimum distance criterion to estimate the life loss; [18]

Manuscript received May 6, 2018; revised September 22, 2018; accepted November 15, 2018. Date of publication November 19, 2018; date of current version January 22, 2019. Paper no. TPWRD-00509-2018. *(Corresponding author: Ming Dong.)*

M. Dong is with ENMAX Power Corporation, Calgary, AB T2G 4S7, Canada (e-mail: Dongming516@gmail.com).

A. B. Nassif is with ATCO Electric, Edmonton, AB T5J 2V6, Canada (e-mail: alexandre.nassif@atcoelectric.com).

B. Li is with Energy Ottawa, Ottawa, ON K1G 3S4, Canada (e-mail: benzheli@energyottawa.com).

Color versions of one or more of the figures in this paper are available online at http://ieeexplore.ieee.org.

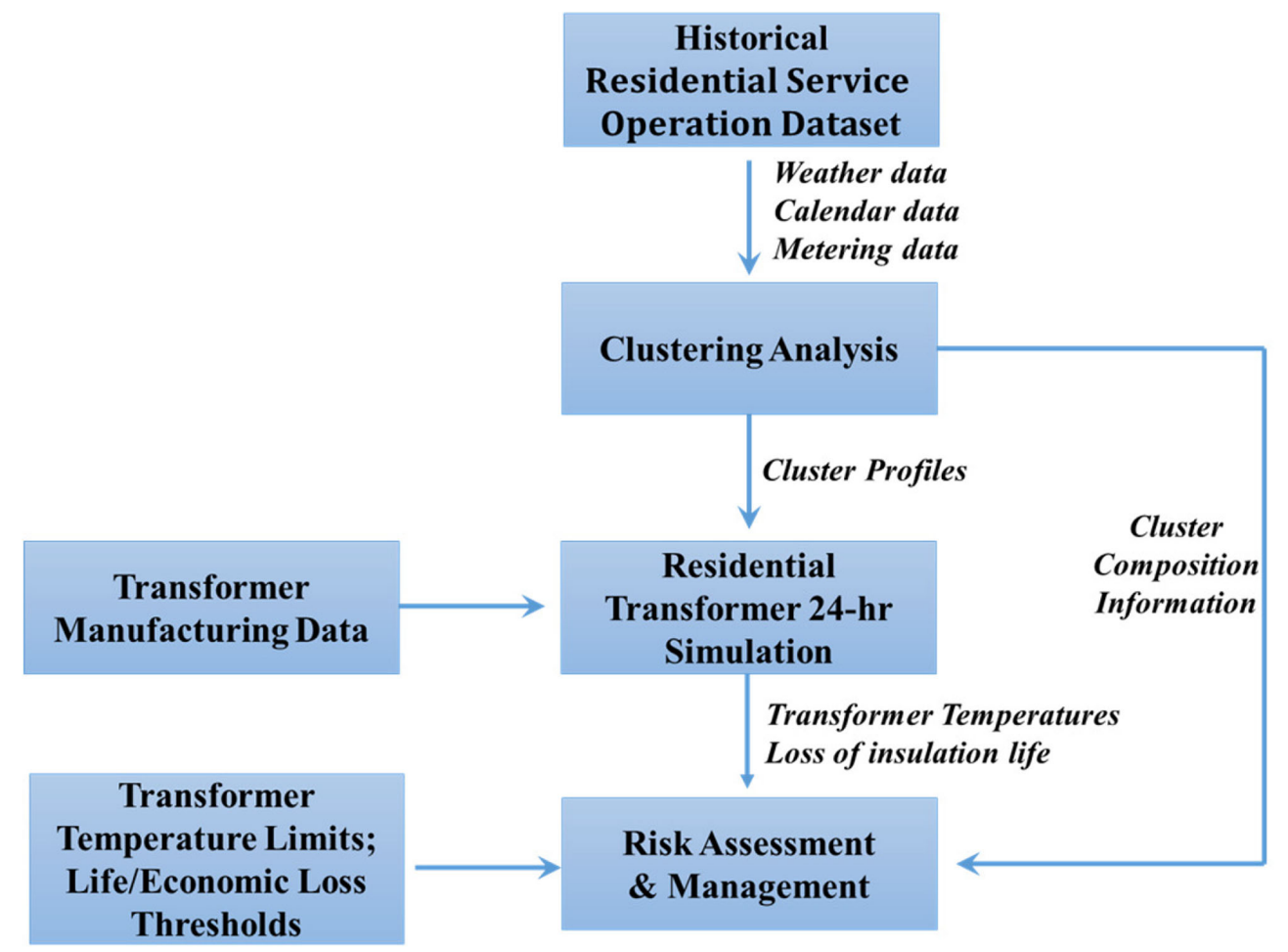

Fig. 1. Flowchart of proposed residential transformer overloading risk assessment.

demonstrated the application of clustering analysis to smart meter data for the extraction of appliance-level loading profiles.

This research proposes a novel statistical data-driven method to assess the overloading risk of oil-immersed residential transformers following the process illustrated in Fig. 1. Currently, Oil-immersed transformers are the most common residential transformers in Canada and US due to its lower cost, lower losses and more compact sizes [19]. Clustering analysis is applied to residential service operation dataset of several recent years in the studied area. The results of clustering analysis have two parts: one part containing the cluster composition information for this area; the other part containing cluster profiles that can be fed into a residential transformer 24-hour simulation model to produce estimated transformer temperatures and insulation life loss. This outcome, combined with cluster composition information, can be used for comprehensive overloading risk assessments as discussed in detail in this paper. The proposed method can quantitatively and statistically assess residential transformer population in one area. Furthermore, proper asset management decisions can be developed according to the assessment results to mitigate transformer overloading risks.

This paper firstly reviews the factors contributing to residential transformer temperature rise and insulation life loss, and then presents an iterative computation model to simulate the 24-hour residential transformer temperature variation and aging behavior based on specific input data. It then takes a deep dive into the residential service operation dataset and discusses the application of clustering analysis and how to feed the analysis results into the established simulation model to derive desired outcomes for overloading risk assessment and management strategies. This method has been applied to an actual Canadian utility company and application examples are presented to illustrate its applicability and usefulness.

## II. Transformer Temperature Rise

IEEE standard C57.91-2011 explains how residential oil-immersed transformer temperature can be affected by transformer loading and ambient temperatures [20], [21]. Temperature variables including transformer top-oil temperatures and winding hot-spot temperatures are functions of 24-hour transformer loading profile and 24-hour ambient temperature profile. With the ambient temperature data and residential service loading data, we can calculate the transformer top-oil temperature and hot-spot temperature by following the steps below.

### A. *Calculate Transformer Top-Oil Temperature*

$$\emptyset_{TO} = \emptyset_A + \Delta\emptyset_{TO} \tag{1}$$

where $\emptyset_{TO}$ is the top-oil temperature, in °C; $\emptyset_A$ is the ambient temperature during the load cycle to be studied, in °C. In our case, this is the hourly ambient temperature in 24 hours during a day; $\Delta\emptyset_{TO}$ is the top-oil rise over ambient temperature, in °C. $\Delta\emptyset_{TO}$ can be further calculated using the formula below:

$$\Delta\emptyset_{TO} = (\Delta\emptyset_{TO,U} - \Delta\emptyset_{TO,i})\,(1 - e^{\frac{24}{\tau_{TO}}}) + \Delta\emptyset_{TO,i} \tag{2}$$

where $\Delta\emptyset_{TO,U}$ is the ultimate top-oil rise over ambient temperature in one hour, in °C; $\Delta\emptyset_{TO,i}$ is the initial top-oil rise over ambient temperature in one hour, in °C; $\tau_{\mathrm{to}}$ is the oil time constant of transformer for any load and for any specific temperature differential between the ultimate top-oil rise and initial top-oil rise. This constant is determined by the weight of core and coil assembly, weight of tank and fittings and liters of transformer oil. It can be derived based on transformer specifications provided by transformer manufacturers.

$\Delta\emptyset_{TO,U}$ is given by the following equation:

$$\Delta\emptyset_{TO,U} = \Delta\emptyset_{TO,R}\left[\frac{(K_U^2 R + 1)}{(R\ +\ 1)}\right]^n \tag{3}$$

where $\Delta\emptyset_{TO,R}$ is the top-oil rise over ambient temperature at rated load on the tap position to be studied, in °C. This is a constant based on the transformer specifications provided by manufactures; $\Delta\emptyset_{TO,U}$ is the ultimate top-oil rise over ambient temperature for load L, in °C; $K_U$ is the ratio of ultimate load to rated load, in per unit; *n* is 0.8 for Oil Natural, Air Natural cooled transformers (most distribution oil-immersed transformers); *R* is the ratio of load loss at rated load to no-load loss, which is also given in the transformer specifications provided by the transformer manufacturer.

It should be noted, in our case, each hour's $\Delta\emptyset_{TO}$ will be calculated based on the initial $\Delta\emptyset_{TO,i}$ which is unknown in the beginning. However, $\Delta\emptyset_{TO,i}$ of hour $k$ is also the $\Delta\emptyset_{TO}$ of hour $k-1$. Therefore, $\Delta\emptyset_{TO,i}$ in the first hour of the day can be initialized to a low temperature number such as 0 °C and then both $\Delta\emptyset_{TO,i}$ and $\Delta\emptyset_{TO}$ of all other 23 hours can be updated. Recurrently, $\Delta\emptyset_{TO}$ of the last hour of the day will be fed back the first hour to replace its initial $\Delta\emptyset_{TO,i}$.This will again trigger the update of the other 23 hours in the day. This computation continues iteratively until all hours reach a steady stage where no significant temperature change can be observed. This stage will indicate the final expected top-oil temperatures of all 24 hours in this daily cycle.

TABLE I
TRANSFORMER TEMPERATURE LIMITS AS PER IEEE STANDARD C57.91-2011

| Temperature Variables | Limit |
|---|---|
| Transformer Top-oil temperature $\emptyset_{TO}$ | 120°C |
| Transformer Hottest-spot temperature $\emptyset_H$ | 200°C |

### *B. Calculate Transformer Winding Hottest-Spot Temperature*

The winding hottest-spot temperature is obtained by solving

$$\begin{cases} \emptyset_H = \emptyset_A + \Delta\emptyset_{TO} + \Delta\emptyset_H \\ \Delta\emptyset_H = (\Delta\emptyset_{H,U} - \Delta\emptyset_{H,i})\left(1 - e^{\frac{24}{\tau_W}}\right) + \Delta\emptyset_{H,i} \end{cases} \quad (4)$$

where $\emptyset_H$ is the winding hottest-spot temperature, in °C; $\Delta\emptyset_H$ is the winding hottest-spot rise over top-oil temperature, in °C; $\Delta\emptyset_{H,U}$ is the ultimate winding hottest-spot rise over top-oil temperature in one hour, in °C; $\Delta\emptyset_{H,i}$ is the initial winding hottest-spot rise over top-oil temperature in one hour, in °C.$\tau_W$ is the winding time constant at hot spot location which can be provided by the transformer manufacturer.

$\Delta\emptyset_{H,U}$ is given by the following equation:

$$\Delta\emptyset_{H,U} = \Delta\emptyset_{H,R} K_u^{2m} \quad (5)$$

where $\Delta\emptyset_{H,R}$ is a constant called hotspot differential and can be provided by the transformer manufacturer; $K_U$ is the ratio of ultimate load L to rated Load, per unit; *m* is 0.8 for Oil Natural, Air Natural cooled transformers.

Similar to the iterative computation process illustrated for the calculation of top-oil temperatures in Section II-A, the 24-hour winding hotspot temperatures can be calculated provided the 24-hour loading and ambient temperature profiles.

### *C. Residential Transformer Temperature Limits Suggested by IEEE Standard C57.91-2011*

IEEE standard C57.91-2011 suggests that the above calculated transformer top-oil temperatures and winding hottest-spot temperatures do not exceed the limits specified in Table I to prevent transformer failures. These limits can be used as safety thresholds for residential transformer loading assessment. Exceeding these limits due to overloading should not be allowed.

## III. RESIDENTIAL TRANSFORMER INSULATION LIFE

Transformer insulation life is also affected by transformer loading and ambient temperature. Experimental evidence indicates that the relation of insulation deterioration to time and temperature follow an adaption of Arrhenius reaction rate theory [21]. IEEE standard C57.91-2011 suggests calculating the hourly aging acceleration factor $F_{AA,n}$ and daily equivalent aging factor $F_{EQA}$ by using the equations below:

$$\begin{cases} F_{AA,n} = e^{\left[\frac{15000}{383} - \frac{15000}{\emptyset_{H,n}+273}\right]} \\ F_{EQA} = \frac{\sum_{n=1}^{24} F_{AA,n}}{24} \end{cases} \quad (6)$$

where $\emptyset_{H,n}$ is the winding hottest-spot temperature in °C for $n_{th}$ hour in 24-hour period, which could be calculated using the methods discussed in Section II.

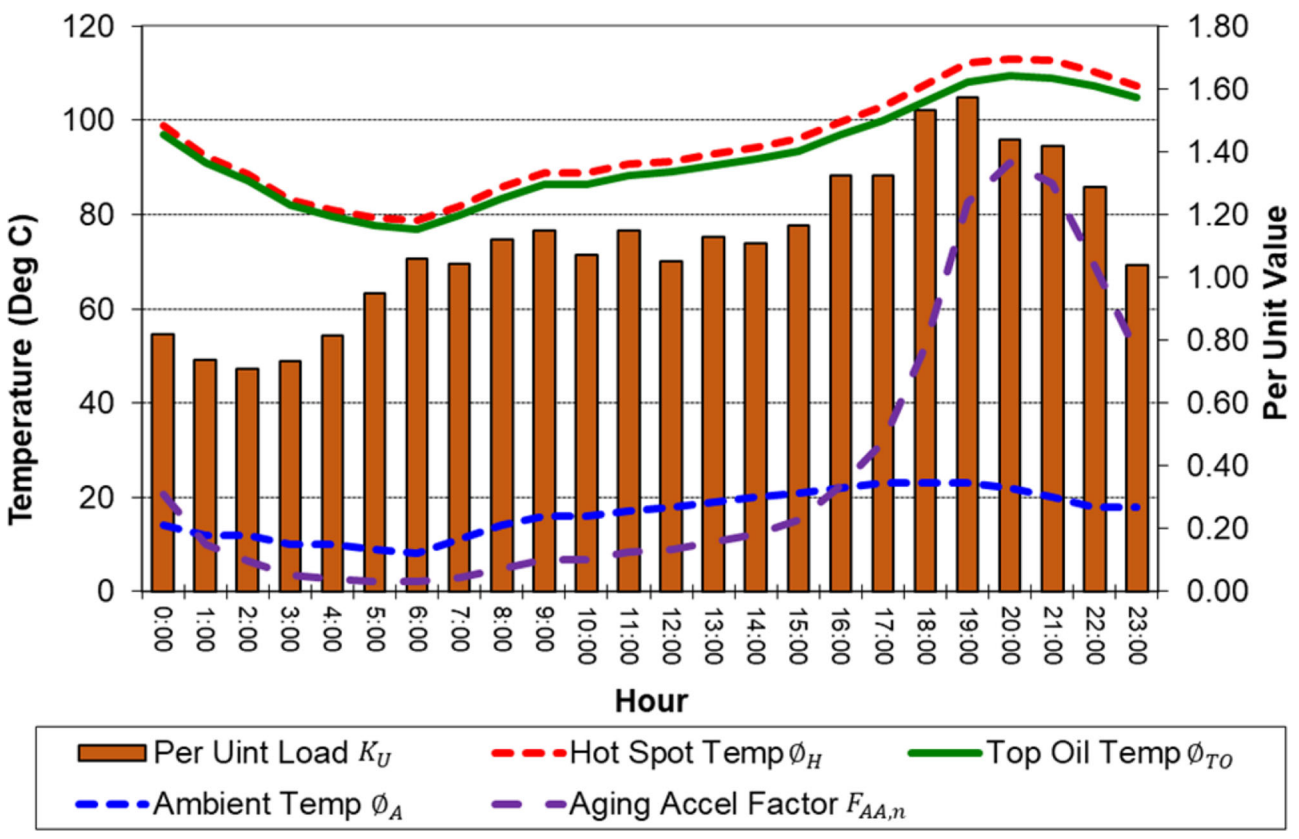

Fig. 2. Example of residential transformer 24-hour simulation results.

Equivalent aging factor $F_{EQA}$. indicates the effective aging time within a 24-hour window. Normal operation expects equivalent aging factor to be one. For instance, if equivalent aging factor is 2, it means the transformer ages by 2 days under normal operating condition in one natural day. Using these equations, we can estimate the loss of transformer insulation life in any given period of time such as one year. This can also be used as a criterion for assessing residential transformer overloading risk. In addition, loss in transformer insulation life can be further converted to monetary loss based on the material and labor costs of purchasing and installing a residential transformer.

## IV. RESIDENTIAL TRANSFORMER SIMULATN

A residential transformer simulation model is established to implement calculations discussed in Section II and III. This computation process uses the following data as input:

- Transformer manufacturer data: weight of core and coil assembly, weight of tank and fittings, liters of transformer oil, top-oil rise over ambient temperature at rated load, ratio of load loss at rated load to no-load loss, winding time constant at hot spot location;
- 24-hour load profile: the per-unit loading value for each hour for a specific transformer. If the number of services connected to the transformer is given, the 24-hour load profile is equal to the diversified service load profile multiplied by the number of services;
- 24-hour ambient temperature in °C.

Through the iterative computation process explained in Section II, this model outputs 24-hour top-oil temperature profile, winding hottest-spot temperature profile and aging accelerated factor profile. An example of the output is shown in Fig. 2.

## V. APPLYING CLUSTERING ANALYSIS TO RESIDENTIAL SERVICE OPERATION DATASET

Applying clustering analysis to historical residential service operation dataset is the key step in the proposed transformer overloading risk assessment method. Clustering is the task of grouping a set of objects in such a way that objects in the same group (cluster) are more similar to each other than to those in other groups (clusters). Clustering is an important unsupervised learning technique that can be used for statistical data analysis [28]. It is applied to the residential service operation dataset in order to:

- evaluate the internal similarities between data points and automatically restructure the dataset to group similar load profiles under similar weather conditions on similar calendar days;
- study the composition of each cluster and understand the risk levels of transformers at different times of year or under different operational conditions;
- extract typical 24-hour load and ambient temperature profiles from each cluster to calculate the corresponding transformer temperatures and insulation life loss to support further risk assessment.

### A. Description of Residential Service Operation Dataset

For a specific area, the residential service operation dataset contains the historical operational condition data and loading data from selected residential metered services. It is defined to consist of three categories of data as follows.

*1) Historical Weather Data:* References [22]–[25] indicate that weather can significantly alter the residential loading behavior. For example, in hot summer days, air conditioning can be heavily used while in cold winter days, electric or gas-fired furnaces will run more frequently in residential houses. Gas furnaces consume much less energy than electric heaters, but still use electricity to constantly circulate hot and cold air. Cold and hot ambient temperatures may also change occupants' behaviors because they may tend to stay home when it is cold or hot outside and this behavior often lead to increased residential loading. 24-hour ambient temperature data is required for transformer temperature and insulation life calculation as explained in Section II and III. This data is often available from government or weather stats websites [15]. Precipitation like rain or snow could also affect people's tendencies to stay indoor or outdoor [26]. In some areas such as Scandinavia, the duration of daylight change significantly between seasons and also affects people's behavior at home. In general, the selection of weather features should be based on the local conditions in the studied area. Also, multiple years of weather data should be used to derive statistically meaningful results. This is because weather data in one area could vary between years [22], [23].

*2) Historical Calendar Data:* Weekdays, weekends as well as statutory holidays could result in very different residential load profiles [22], [25], [26]. For example, in some areas, the average daytime consumption during weekdays is typically lower than that in the weekends, and in the evening the average consumption is somewhat higher than that in the weekend evenings [22].

*3) Historical Meter Data:* As discussed above, weather and calendar data will significantly affect average residential load behavior. However, load behavior could also vary under similar weather conditions and on similar calendar days. For example, one day in September may have similar weather conditions as one day in August and they are both workdays, however the loading on the day in August could be much higher than the day in September because in August children could be taking summer break and spending more time at home, while in September they could be going to school [26]. Other drivers for change include the adoption of energy efficient appliances such as LED lights [26], the adoption of new appliances, such as EV chargers [7], over time. This behavioral difference can be reflected directly on residential metering data.

Interval meters or smart meters are measuring the continuous power demand of connected residential services and can provide 24-hour service load profiles. This can be used to produce transformer load profile as explained in Section IV. Some considerations to be accounted for are:

- It is not necessary and could be computationally prohibitive to utilize every residential house's metering data in the studied area. Depending on the availability of interval meters or smart meters a utility company has installed and the size of the studied area, the required number of meters could vary;
- To be statistically representative, it is recommended to collect metering data from different residential communities across the studied area instead of being concentrated in one or two residential communities. This is to ensure the extraction of statistically unbiased residential load behaviors;
- The metering data should span over multiple years. This is to match the requirement of using multi-year ambient temperature data.
- There are areas where no interval meters or smart meters are installed. Traditional revenue meters only record cumulative energy consumption change between readings. The energy consumption data can be converted to average loading data for a certain period of time but cannot provide continuous 24-hour load profiles. For transformers in an area like this, this section proposes a Euclidian distance based method in the end to leverage the assessment results in another similar area that has continuous load profiles to estimate the transformers in the target area.

### B. Basics of K-Means Clustering

K-Means clustering is selected as the unsupervised learning method to process the above residential service operation dataset. K-Means method is a very popular unsupervised clustering method for dealing with large datasets with great efficiency and simplicity [28]–[31]. It only requires one parameter $k$ which is defined as the expected number of clusters produced from the clustering process. The mathematic description of K-Means clustering is stated as below.

Given a set of observations $(X_1,\ X_2,\ldots,X_n)$, where each observation is a d-dimensional real vector, K-Means clustering aims to partition the observations into $k$ ($\leq n$) sets $S = \{S_1,\ S_2,\ldots,S_k\}$ so as to minimize the within-cluster sum of squares (i.e., variance). Formally, the objective is to find:

$$\arg_s^{\min}\sum_{i=1}^{k}\sum_{\mathrm{x}\in S_i}\left\|\mathrm{x}-\mu_i\right\|^2 = \arg_s^{\min}\sum_{i=1}^{k}|S_i|\,\mathrm{Var}\,S_i \qquad (7)$$

where $\mu_i$ is the mean of $S_i$ [27].

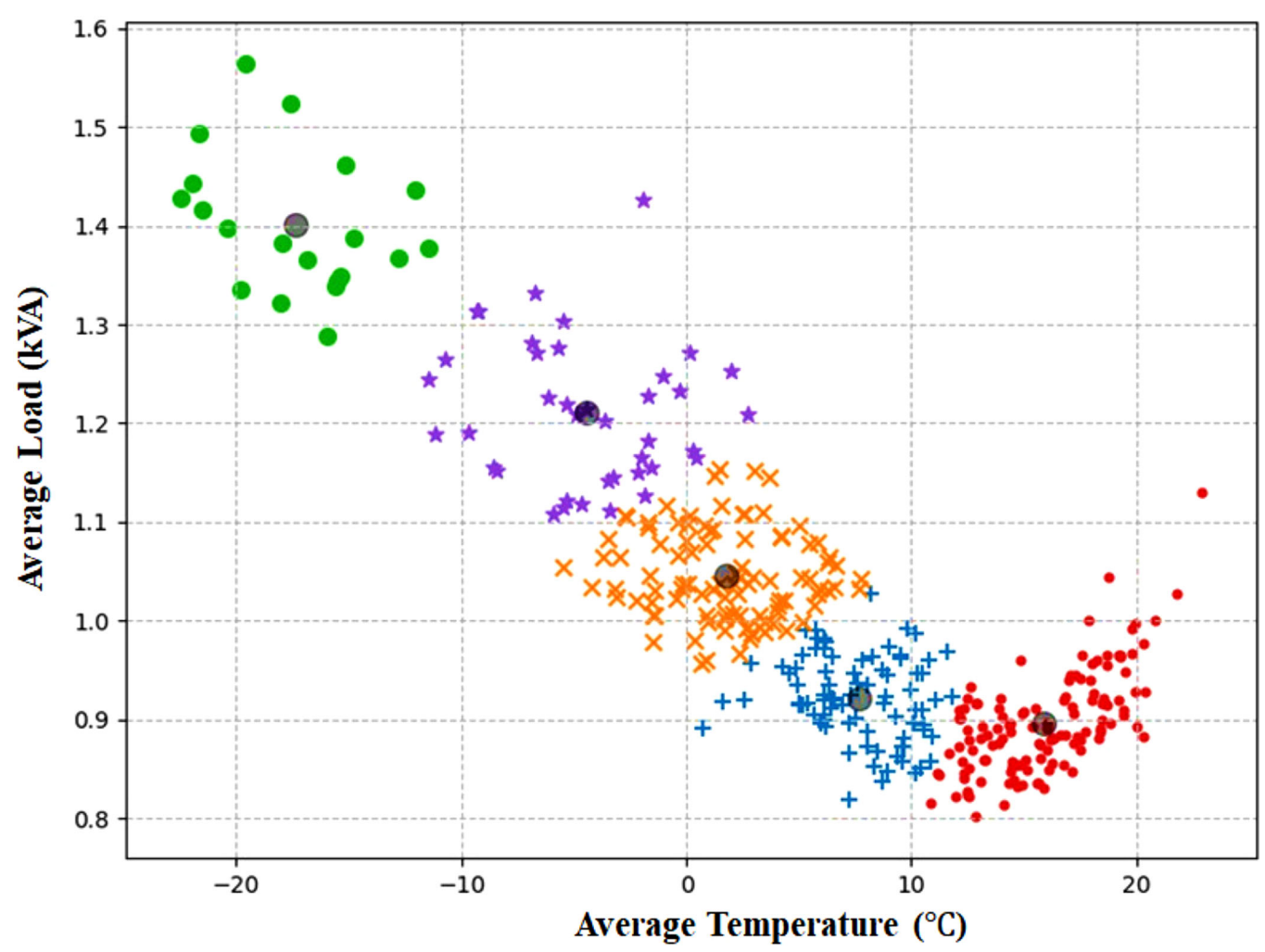

Fig. 3. Example of clustering results using two features $T_{avg}$ and $L_{avg}$ (k = 5).

For the residential service operation dataset, an example of a feature vector (*FV*) for a historical day is given as below.

$$FV = \left(T_{max}, T_{min}, T_{avg},\ L_{max}, L_{min}, L_{avg},\ C_{rain},\ C_{weekday}\right) \tag{8}$$

In this example, for a given date, the maximum, minimum and average daily ambient temperatures and loading levels are recorded into the feature vector *FV*. In addition, *FV* has two categorical features to indicate how rainy the day is and if it is a weekday. In practice, the selection of features depends on feature data availability, the variation of feature data and the sensitivity of feature data to the clustering results.

The standard steps of K-Means clustering are:

- *Step 1:* Initialize *k* centroids randomly within the data domain;
- *Step 2:* Associate all data points to their nearest centroids. This will create *k* data clusters. Each cluster contains the associated data points as its members;
- *Step 3:* Update the centroid of each cluster using all members in the cluster;
- *Step 4:* Repeat Step 2 and Step 3 until convergence has been reached.

For the purpose of concept illustration, the clustering result of using only features $T_{avg}$ and $L_{avg}$ for 365 records is plotted in Fig. 3. As can be seen, all 365 records were grouped into 5 clusters.

### C. Modified K-Means Clustering to Deal With Categorical Data

Residential service operation dataset could include categorical data to describe weather and calendar conditions. Orderly categorical data such as moderate, medium, severe can be converted into numerical values using (9):

$$x = \frac{i - 1/2}{N},\ i = 1, 2 \ldots N \tag{9}$$

where $N$ is the total orderly statuses, $i$ is the order of the status [28].

Another type of categorical data is simply indicating unordered statuses such as yes or no. This would require the modification of dissimilarity calculation in the standard K-means algorithm, which is based on Euclidian distance. The modified dissimilarity calculation between condition feature vector *X* and *Y* would look like:

$$\begin{cases} d\left(X, Y\right) = \sum_{j=1}^{p} \left(x_j - y_j\right)^2 + \sum_{j=p+1}^{m} \delta\left(x_j, y_j\right) \\ \delta\left(x_j, y_j\right) = \begin{cases} 0, \left(x_j = y_j\right) \\ 1, \left(x_j \neq y_j\right) \end{cases} \end{cases} \tag{10}$$

where $x_1$ *to* $x_p$ and $y_1$ *to* $y_p$ are all numerical features and $x_{p+1}$ *to* $x_m$ and $y_{p+1}$ *to* $y_m$ are unordered categorical features. If two unordered categorical features match, the distance of categorical features is 0; otherwise, it is 1 [31].

### D. Normalization and Weightings of Condition Features

To effectively apply K-means clustering, all features should be normalized to a numerical range (e.g., [0, 1]). This is because the raw residential service operation data use different units and the magnitude differences of different features can be quite large. There are many ways of normalizing raw features, for example, the Min-Max normalization can be described as below:

$$x_{norm} = \frac{x_{raw} - Min}{Max - Min} \tag{11}$$

where for a give dataset, *Max* is the maximum value observed in feature *j; Min* is the minimum value observed in feature *j* [28]–[30].

In addition to normalization, different weighting factors can be assigned to the dissimilarity formula when applying K-Means clustering:

$$d\left(X, Y\right) = \sum_{j=1}^{p} w_j \left(x_j - y_j\right)^2 + \sum_{j=p+1}^{m} w_j \delta\left(x_j, y_j\right) \tag{12}$$

where $w_j$ is the empirical weighting factor for feature *j.* For example, in places where temperatures do not vary much or no heating or cooling devices are used in residential units, the weightings on temperature features can be lowered accordingly.

### E. Clustering Composition

Following the above steps, with the produced clustering results, utility asset management engineers can:

- examine the data points in each cluster to understand the distribution of dates inside each cluster;
- calculate the centroid of clusters. The centroid of a cluster is the averaged feature vector from all members in the cluster and is an indication of the average loading behavior and operation conditions of all members in the cluster.

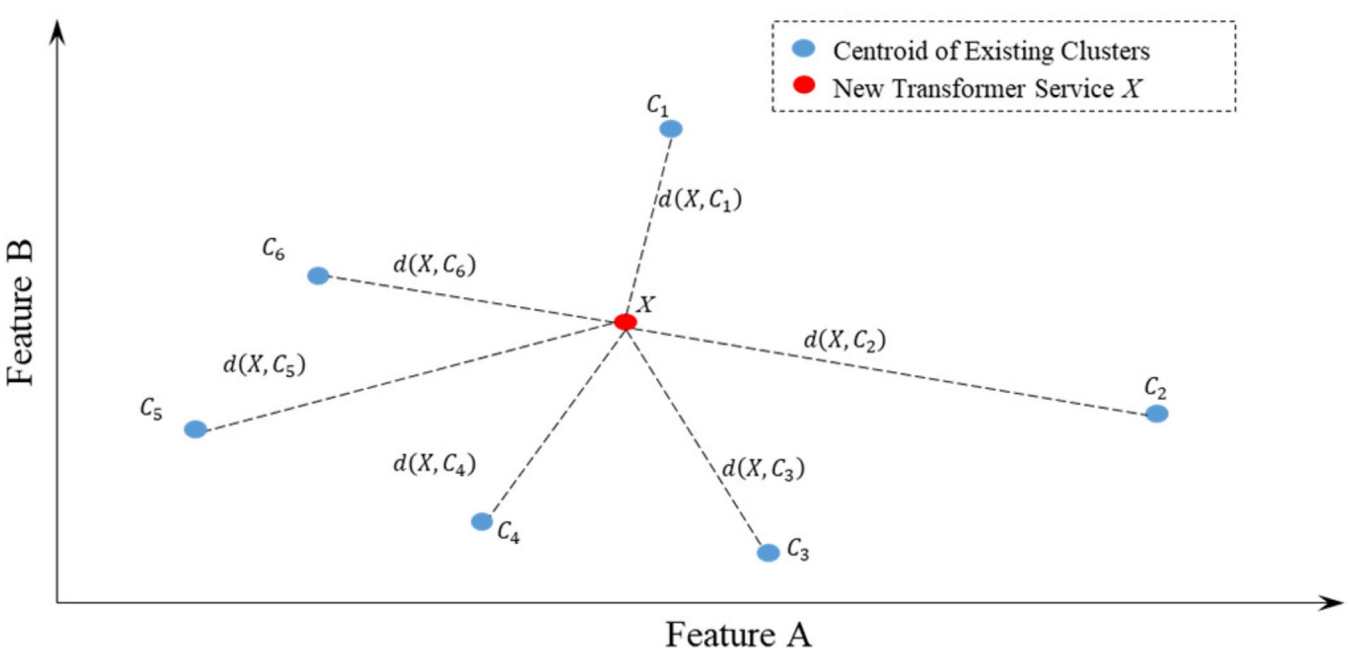

Fig. 4. The proposed method of leveraging existing cluster centroids to estimate new transformer service *X* based on $d(X, C_i)$.

### *F. Cluster Profiles and Residential Transformer Simulation*

For each cluster, the 24-hour load and ambient temperature profiles representing the common behaviors within the cluster can be extracted and these profiles are named as cluster profiles. The cluster profiles can be calculated by averaging the profiles of all members within the cluster or averaging the profiles of a number of randomly selected members within the cluster. As shown in Fig. 1, a cluster profile can be scaled up by the number of services connected to the transformer. The scaled profile is then fed into the residential transformer 24-hour simulation model established in Section IV to estimate transformer temperature rises and life losses. Combined with the cluster composition results, utility asset management engineers can:

- determine the transformer loading thresholds throughout a year;
- rank the clusters based on their impact on transformers;
- given the number of residential services *n*, calculate the transformer temperatures and life loss for each cluster.
- given the number of residential services *n*, estimate temperatures for a new transformer that has *n* services and an average service feature vector *X*. The equation is developed based on the similarities of *X* with respect to all cluster centroids measured by Euclidian distances:

$$T_{x,n} = \sum_{i=1}^{k} T_{i,n} \left( \frac{1/d\left(X, C_i\right)}{\sum_{i=1}^{k} 1/d\left(X, C_i\right)} \right) \tag{13}$$

where *k* is the total number of existing clusters; $T_{i,n}$ is the transformer temperature under an existing cluster *i* with *n* residential services; $d(X, C_i)$ is the Euclidian distance of feature vector *X* to existing cluster centroid $C_i$.

The meaning of (13) is explained with respect to Fig. 4. Intuitively, if the given feature vector *X* is very close to an existing cluster centroid $C_i$, *X*'s load profile and ambient temperature profile should also be very close to $C_i$'s load profile and temperature profile. This implies that the resulted transformer's temperature should be very close to $T_{i,n}$.In this case, Euclidian distance $d(X, C_i)$ will be close to zero and (13) would end up to be $T_{x,n} \approx T_{i,n}$. In another case where *X* is surrounded by all clusters, $T_{x,n}$ will be the average weighted by the similarities between *X* and all cluster centroids. This method could provide reasonable approximation if *X* is not far away from all clusters. If *X* is far away from all clusters, this means *X* is not part of the established learning model and therefore should not rely on this method to estimate. Geographically, this suggests transformer *X* should only be in the studied area or areas which have similar weather and electricity use characteristics compared to the originally studied area. For a provincial or national utility company owning multiple similar geographic areas, this is often the case. In contrast to this, a cluster model developed based on tropical areas should not be applied to transformers in cold areas; a cluster model characterizing electricity uses in house-concentrated areas should not be applied to condo-concentrated areas.

- Similar to estimating transformer temperatures, given a number of residential services *n*, the insulation life loss of a new transformer *X* can be estimated too. Equation (13) now becomes:

$$l_{x,n} = \sum_{i=1}^{k} l_{i,n} \frac{1/d\left(X, C_i\right)}{\sum_{i=1}^{k} 1/d\left(X, C_i\right)}) \tag{14}$$

where *k* is the total number of existing clusters; $l_{i,n}$ is the transformer insulation life loss under cluster *i* with *n* residential services; $d(X, C_i)$ is the Euclidian distance of feature vector *X* to existing centroid $C_i$.

## VI. Application Example

The proposed method was recently applied to a utility company in West Canada. Multiple application examples are provided in this section to explain the process and outcomes of the proposed method. The residential service operation dataset used in this section includes:

- Local hourly ambient temperature data in a City from 2014 to 2016;
- Calendar data from 2014 to 2016;
- Hourly loading data from 125 residential metered services across the city from 2014 to 2016.

### *A. Determine the Members and Centroids of Clusters*

First, the above described dataset was processed to extract 5 features for all 125 services from 2014 to 2016 on a daily basis. These 5 features are maximum daily ambient temperature, minimum daily ambient temperature, average daily ambient temperature and the weekday/weekend categorical feature. The total number of records used in this analysis is 137,000 (1096 days times 125 services). Then the first 4 non-categorical features were normalized using (11). In the end these features were taken into the steps described in Section V-B for K-means clustering. It should be noted due to the presence of one categorical feature, modified Euclidean distance (10) is chosen for calculating dissimilarities between data points. Table II summarizes the produced cluster composition information. The total of 137,000 data points were automatically grouped into 10 clusters based on their similarities defined by the 5 features. The centroid values of clusters are also listed in Table II to indicate the common characteristics of each cluster's members. In practice, the number of clusters can be estimated by human experts with some prior knowledge on the intrinsic or desired structure of the input data [32]. In this example, the whole year temperature is estimated to have 5 distinct levels: extremely low, low, medium,

TABLE II
CLUSTER COMPOSITION INFORMATION

| Cluster ID | Number of Members | $T_{max}$ (°C) | $T_{min}$ (°C) | $T_{avg}$ (°C) | $L_{avg}$ (kVA) | Weekday (Y/N) |
|---|---|---|---|---|---|---|
| 1 | 139 | 5.14 | -4.57 | 0.47 | 1.66 | Y |
| 2 | 138 | 20.88 | 7.93 | 14.50 | 1.40 | N |
| 3 | 176 | 18.08 | 5.04 | 11.54 | 1.33 | Y |
| 4 | 58 | -11.12 | -20.58 | -15.78 | 2.00 | Y |
| 5 | 46 | 0.35 | -11.35 | -5.59 | 1.95 | N |
| 6 | 168 | 24.92 | 11.00 | 18.22 | 1.37 | Y |
| 7 | 107 | 8.64 | -2.32 | 3.13 | 1.63 | N |
| 8 | 155 | 12.32 | -0.75 | 5.67 | 1.46 | Y |
| 9 | 23 | -14.06 | -22.83 | -18.67 | 2.15 | N |
| 10 | 87 | -0.90 | -11.69 | -6.17 | 1.84 | Y |

TABLE III
TRANSFORMER THRESHOLD AND CLUSTER IMPACT RANKING BY MAXIMUM 24-HOUR PEAK LOADING (p.u.)

| Cluster ID | Maximum 24-hr average loading (p.u.) | Maximum 24-hr peak loading (p.u.) | Impact Ranking |
|---|---|---|---|
| 1 | 1.80 | 2.52 | 6 |
| 2 | 1.70 | 2.20 | 2 |
| 3 | 1.72 | 2.29 | 3 |
| 4 | 2.00 | 2.75 | 10 |
| 5 | 1.87 | 2.58 | 7 |
| 6 | 1.62 | 2.19 | 1 |
| 7 | 1.82 | 2.40 | 5 |
| 8 | 1.79 | 2.39 | 4 |
| 9 | 2.00 | 2.73 | 9 |
| 10 | 1.87 | 2.65 | 8 |

TABLE IV
NUMBER OF DAYS IN 12 MONTHS IN EACH CLUSTER

| Mth | Imp 1 | Imp 2 | Imp 3 | Imp 4 | Imp 5 | Imp 6 | Imp 7 | Imp 8 | Imp 9 | Imp 10 |
|---|---|---|---|---|---|---|---|---|---|---|
| Jan | 0 | 0 | 0 | 7 | 9 | 26 | 10 | 22 | 8 | 11 |
| Feb | 0 | 0 | 0 | 11 | 12 | 22 | 7 | 13 | 5 | 15 |
| Mar | 0 | 1 | 5 | 24 | 19 | 21 | 5 | 10 | 2 | 6 |
| Apr | 2 | 7 | 22 | 33 | 18 | 7 | 0 | 1 | 0 | 0 |
| May | 11 | 17 | 32 | 20 | 11 | 2 | 0 | 0 | 0 | 0 |
| June | 33 | 25 | 32 | 0 | 0 | 0 | 0 | 0 | 0 | 0 |
| July | 62 | 26 | 5 | 0 | 0 | 0 | 0 | 0 | 0 | 0 |
| Aug | 47 | 28 | 17 | 1 | 0 | 0 | 0 | 0 | 0 | 0 |
| Sep | 12 | 20 | 42 | 10 | 4 | 2 | 0 | 0 | 0 | 0 |
| Oct | 1 | 11 | 18 | 35 | 16 | 12 | 0 | 0 | 0 | 0 |
| Nov | 0 | 2 | 3 | 13 | 16 | 28 | 7 | 13 | 2 | 6 |
| Dec | 0 | 0 | 0 | 1 | 2 | 19 | 17 | 28 | 6 | 20 |
| Sum | 168 | 137 | 176 | 155 | 107 | 139 | 46 | 87 | 23 | 58 |

*Imp N represents the cluster with impact level N as per Table III.

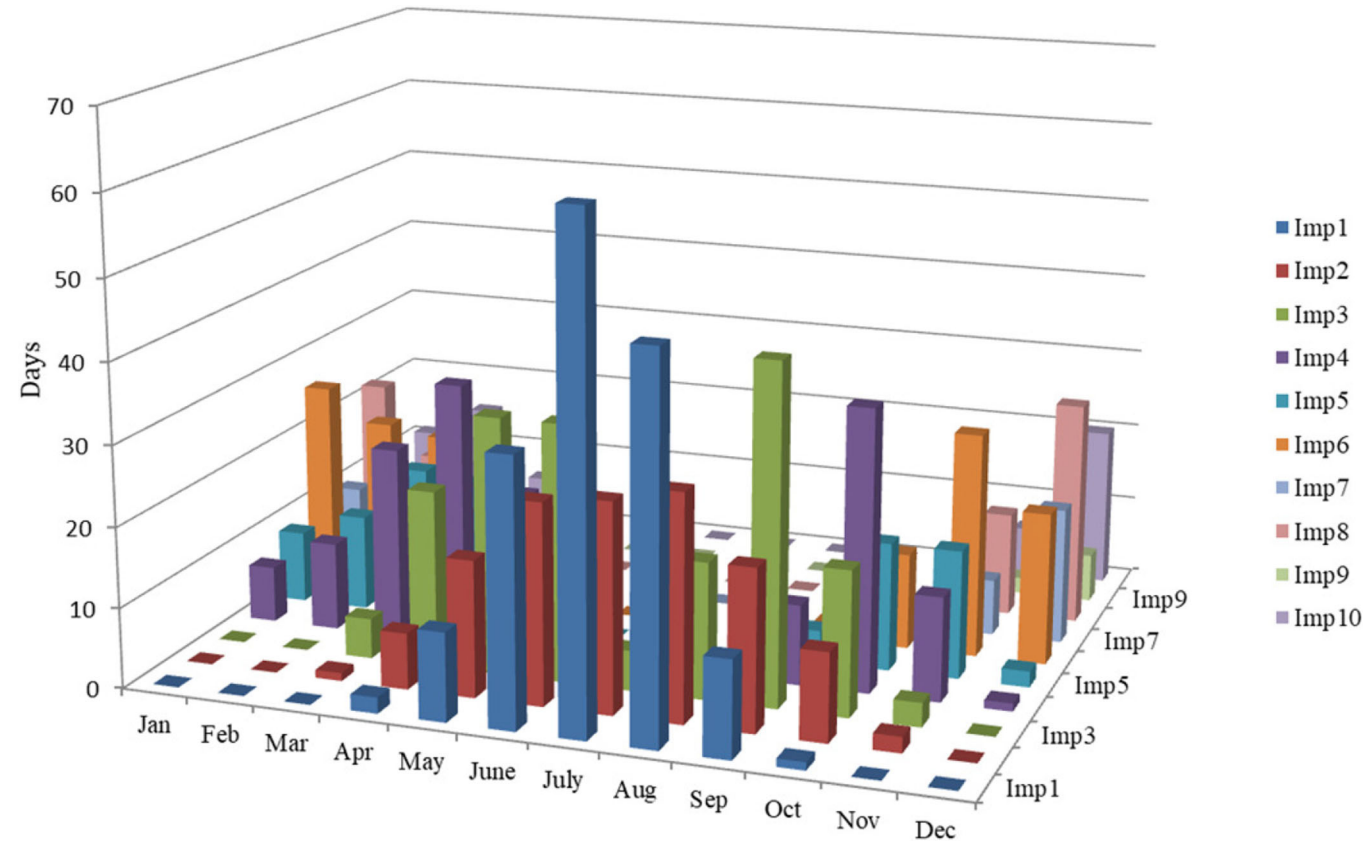

Fig. 5. Number of days in 12 months in each cluster.

high and extremely high. This multiplied by the binary weekday/weekend attribute yields 10 clusters in total. However, in another geographic area where more weather conditions could affect residential loading behavior, the required number of clusters may need to increase to reflect the nuances. For example, in some places, rains are frequent throughout a whole year. It may be wise to consider additional number of clusters to reflect low rain, medium rain and heavy rain scenarios. Another way to determine the number of clusters is to initialize a number and increase it gradually to an extent that no significant change is observed from the assessment results. Depending on the size of the dataset, this method can be very computational intensive.

## *B. Determine the Transformer Loading Thresholds and Rank Cluster Impact*

The cluster profiles were then extracted and taken into a 25 kVA residential transformer 24-hour simulation model established by following Section IV. The loading profile was scaled up gradually until either the top-oil temperature limit or winding hottest-spot temperature was reached. In this example, top-oil temperature limit was used as the overloading criterion. The corresponding transformer average loading and peak loading in per unit were recorded. The results are shown in Table III. Each cluster's impact level was ranked by the magnitude of peak loading. It indicates the sensitivity of transformer temperature and life loss to loading. For instance, it is found out Cluster 6 can only go up to 2.19 per unit. With reference to Table II, it is found that this cluster represents hot days because all the temperature feature values are high; the impact on transformer temperature is low for clusters 9 and 4 because of their low ambient temperatures. Between these two, Cluster 4's impact is further lower than Cluster 9 because all its members are weekdays which draw a lower average loading than that of Cluster 9 during weekends. The results in Table III also suggest that throughout a year, the minimum allowed daily peak loading is 2.19 p.u. The utility company can therefore use this value as a conservative loading threshold for residential transformers in the City. If a residential transformer's average loading frequently exceeds 2.19 p.u., an alarm can be triggered for further investigation.

## *C. Calculate the Number of Days in Each Month in Each Cluster*

Based on the impact ranking of Table III, we can further examine the dates of the members inside each cluster and the results are summarized in Table IV and Fig. 5. A clear pattern was revealed in this study: high impact level clusters have large concentrations in summer months; the impact level decreases as the dates move away from hottest summer months. For instance, Cluster 7 and 8 have almost zero members from June to August; the lowest impact-level clusters came from winter months (December to March). This suggests that the combined effect of loading and ambient temperature in winter months is found to be less harmful to the transformers. On the contrary, the summer months June, July and August are statistically more harmful.

The numbers of days were also summed for each impact level. It shows that high-impact days are significantly more than the low-impact days in the studied area. For example, Imp 1 to Imp 3 (Clusters 6, 2 and 3) account for 44% of the total days. If the sum of days is used as the weighting factor, the weighted average impact level is close to 4. Imp 4 is corresponding to Cluster 8 as per Table III. This suggests if no other information is provided, as a general rule of thumb, the peak loading of residential transformers in this area should be kept below 2.39 p.u.

More granular monthly or seasonal assessment can also be made from this type of analysis based on the clustering results to produce proper risk management actions. Although monitoring transformer loading individually is not very practical, almost all utility companies have the capability of monitoring feeder-level loading through SCADA system nowadays [33]. For a residential feeder, based on the feeder-level loading and the number of residential transformers connected to the feeder, the average residential transformer loading could be estimated. As shown in Fig. 4, since most of the days in June, July and August belong to Cluster 6 with the highest impact level (i.e., Imp1), the average peak loading threshold should not exceed 2.19 p.u. If 2.19 p.u is frequently exceeded, this is an indication that statistically the transformers on this feeder will be prone to fail due to overloading. Proactive measures can be taken for this feeder. For example, transformers on this feeder may need to be upgraded to larger sizes such as from 25 kVA to 37.5 kVA. The upgrades can be prioritized to start from transformers with more service connections until the observed average transformer loading gets reduced to below 2.19 p.u. Another proactive measure can be installing new transformers adjacent to existing transformers to share the service connections. This will also reduce the average transformer loading and the potential failure risk.

Alternatively, instead of investing capital to upgrade or install transformers on the concerning feeder, the company could also consider incenting the residential customers on this feeder to reduce load especially during peak hours. This can be realized by implementing time-of-use rates in different months and different time of a week [34]–[36]. For example, it is found in Table II that Cluster 6 is weekday. Therefore, the rates during the weekday peak hours in June, July and August can be set higher to encourage spontaneous peak shaving behaviours. Furthermore, the utility company could consider incenting customers on this feeder to install more DERs such as roof-top PV panels to offset the demand and reduce the loading on their connected residential transformers.

Similarly, as show in Fig. 5, Cluster 10 with Impact-level 8 is the most dominant cluster (with tallest bar) in December. Proactive measures can be designed to ensure the average residential transformer peak loading on a feeder does not exceed 2.65 p.u in December, as specified in Table III.

### D. *Determine the Maximum Allowed Number of Services Per Residential Transformer by Temperatures*

Assuming *N* services are connected to a transformer, the transformer's load profile can be calculated as the cluster load profile multiplied by service number *N*. Again, using the residential transformer 24-hour simulation model, the maximum top-oil temperatures or winding hottest-spot temperatures can be calculated. A table like Table V can be developed to test different service numbers. It shows the maximum top-oil temperatures resulted from different service numbers. Since 120 °C is the suggested operating limit for top-oil temperature, it is clear that statistically the maximum number of services should not exceed 23 for the simulated 25 kVA transformer model in the City. This finding can be implemented as a transformer installation and design requirement to prevent new transformers from connecting with more than 23 services. This way, harmful overloading level can be avoided in the first place. This can also be used to identify existing overloaded transformers that have more than 23 services in the population. As discussed in Section VI-C, these transformers can be either upgraded or equipped with additional transformers in adjacency so that their loadings can get down below the thresholds.

TABLE V
NUMBER OF SERVICES N AND TRANSFORMER MAXIMUM TOP-OIL TEMPERATURE (°C) FOR 25 kVA TRANSFORMER

| Cluster ID | N =23 | N =24 | N =25 | N =26 | N =27 | N =28 | N =29 | N =30 |
|---|---|---|---|---|---|---|---|---|
| 1 | 93 | 99 | 106 | 112 | 119 | **126** | **133** | **140** |
| 2 | 85 | 89 | 94 | 99 | 104 | 109 | 114 | 119 |
| 3 | 77 | 81 | 85 | 90 | 94 | 99 | 104 | 109 |
| 4 | 104 | 113 | **121** | **130** | **138** | **147** | **157** | **166** |
| 5 | 113 | **121** | **129** | **138** | **146** | **155** | **164** | **173** |
| 6 | 88 | 93 | 98 | 103 | 108 | 113 | 118 | **123** |
| 7 | 90 | 96 | 102 | 108 | 114 | **121** | **127** | **134** |
| 8 | 80 | 85 | 90 | 95 | 100 | 105 | 111 | 117 |
| 9 | 116 | **125** | **135** | **144** | **154** | **164** | **175** | **185** |
| 10 | 102 | 110 | 117 | **125** | **133** | **141** | **149** | **158** |
| Max | 116 | **125** | **135** | **144** | **154** | **164** | **175** | **185** |

### E. *Determine the Maximum Allowed Number of Services Per Residential Transformer by Loss of Life*

Similar to the study in Section VI-D, transformer loss of life can be calculated for different numbers of services connected to one transformer. The normal life of a distribution transformer is 20.55 years according to [20]. A utility company can determine a desired limit for the average annual loss of life such as 400 days; it can further set a desired economic threshold based on the purchase and installation cost of a new transformer. In this case study, the transformer purchase and installation cost is assumed to be $5,000 and the company decided the maximum acceptable economic loss per transformer is $500/year. Equivalent economic loss for each service number is calculated using the formula below:

$$EL = \frac{L_A}{7500} \times C \tag{15}$$

Where $L_A$ is the average annual loss of life calculated in Table VI; $C$ is the cost of purchasing and installing a new transformer, in this case assumed to be $5,000. 7500 is the normal life expectancy in days required by standard [20] (equivalent to 20.55 years) in a constant 30 °C ambient temperature with a continuous loading at transformer's rated kVA. In practice, both the temperature and loading level fluctuate as shown in Fig. 2 and the corresponding aging process will be almost

TABLE VI
NUMBER OF SERVICES N AND TRANSFORMER LIFE LOSS FOR 25 kVA TRANSFORMER

| Cluster ID | N=19 | N=20 | N=21 | N=22 | N=23 | # of Days |
|---|---|---|---|---|---|---|
| 1 | 0.1 | 0.3 | 0.7 | 1.6 | 3.7 | 139 |
| 2 | 0.1 | 0.1 | 0.2 | 0.4 | 0.7 | 138 |
| 3 | 0.0 | 0.0 | 0.1 | 0.1 | 0.2 | 176 |
| 4 | 0.5 | 1.6 | 4.3 | 11.8 | 31.0 | 58 |
| 5 | 1.2 | 3.3 | 8.7 | 22.3 | 56.0 | 46 |
| 6 | 0.1 | 0.1 | 3.5 | 6.5 | 12.2 | 168 |
| 7 | 0.1 | 0.2 | 0.5 | 1.1 | 2.5 | 107 |
| 8 | 0.0 | 0.1 | 0.1 | 0.2 | 0.4 | 155 |
| 9 | 1.9 | 5.6 | 16.3 | 45.8 | 125.0 | 23 |
| 10 | 0.4 | 1.0 | 2.7 | 6.8 | 16.8 | 87 |
| 3-Year Total Loss of Life (Days) | 213.2 | 567.1 | 2051.9 | 4891.1 | 11735.8 | - |
| Average annual Loss of Life (Days) | 71.1 | 189.0 | 684.0 | 1630.4 | 3911.9 | - |
| Economic Loss ($/year) | 47.4 | 126.0 | 456.0 | **1086.9** | **2608.0** | - |

TABLE VII
ESTIMATED MAXIMUM TOP-OIL TEMPERATURE FOR TRANSFORMER X

| Day | $T_{max}$ (°C) | $T_{min}$ (°C) | $T_{avg}$ (°C) | $L_{avg}$ (kVA) | Weekday (Y/N) | Peak Top-oil Temperature (°C) |
|---|---|---|---|---|---|---|
| 1 | 21.53 | 8.12 | 14.20 | 0.97 | Y | 87.4 |
| 2 | 22.73 | 12.62 | 14.62 | 0.98 | Y | 88.4 |
| 3 | 20.12 | 9.41 | 13.39 | 1.06 | Y | 88.9 |
| 4 | 18.61 | 10.78 | 16.55 | 0.97 | Y | 87.9 |
| 5 | 18.48 | 11.80 | 14.77 | 1.07 | Y | 89.2 |
| 6 | 20.04 | 11.26 | 16.52 | 1.34 | N | 93.5 |
| 7 | 21.72 | 8.18 | 17.58 | 1.23 | N | 92.8 |

always accelerated or decelerated, which would result in a shorter or longer lifespan.

Since the maximum acceptable economic loss per transformer is set to be $500/year, from Table VI the corresponding maximum allowed number of services should be 21 ($456.0/year). Statistically, when the number of services is less than 21, the average annual loss of life will be less than 680.4 days and the equivalent economic loss will be less than $456.0/year. Again, this can be implemented as a transformer installation and design requirement for this area or vice versa, used to identify existing overloaded transformers supplying more than 21 services and mitigate them proactively, as discussed previously.

### *F. Estimate Transformer Temperatures and Loss of Life for Transformers in a Different Area*

As discussed in Section V, (13) and (14) can be used to estimate the maximum top-oil temperatures, winding hottest-spot temperatures and insulation life loss of a transformer located in a different area that shares similar weather and electricity use characteristics to this city. The established clusters in Section VI-A can be leveraged in this application. An example is shown in Table VII. First, the maximum ambient temperature, minimum ambient temperature and average ambient temperature during seven days in this new area were collected. The service revenue meters under transformer *X* are not equipped with interval meters and cannot therefore acquire or record continuous loading data. However, almost all residential revenue meters acquire and record the cumulative energy consumption. After each day, the increased energy consumption can be converted to average service loading (i.e. $L_{avg}$) using the formula below:

$$L_{avg} = \frac{\sum_i^n E_i}{24n} \tag{16}$$

where $E_i$ is the daily energy consumption in kWH recorded from each service revenue meter; $n$ is the number of service connections under transformer *X*.

Taking one step further, multiple transformers under a wider timespan in this area can be estimated in the same way, for example for a month or a year. The estimated maximum top-oil temperatures can be compared to the transformer limits in Table II to identify the overloaded ones. Similar to estimating transformer temperatures, following (14), the loss of life for transformer *X* can be estimated too.

After repeating the above process for more transformer samples in this new area, a statistical view of the residential transformer population in this area can be developed. This type of analysis was not possible without using the proposed data-driven method because the services in this area do not have interval meters or smart meters. As a result, there is no continuous measurement data to support the transformer temperature and aging simulation as discussed in Section IV. Now the proposed method leveraged the clustering results generated in a different area for this area based on readily available data such as ambient temperature and average service loading. Ultimately, risk management measures can be developed for this area to either upgrade or replace overloaded transformers to maintain the transformer temperatures and losses of life at a desired level.

## VII. CONCLUSION

This paper presents a data-driven method for residential transformer overloading risk assessment. Compared to previous works, this method has the following advantages:

- It is reliant on residential service operation data and does not require transformer health or load monitoring devices;
- It provides statistical insights for the overloading risk of residential transformer population in one area;
- Many types of assessment results can be produced and converted to effective risk management measures.

The proposed method was applied to a utility company in West Canada and provided great value to its asset management engineers. Detailed application examples and discussions are given to demonstrate the practical application of this method.

**Ming Dong** (S’08–M’13–SM’18) received the doctoral degree from the University of Alberta, Edmonton, AB, Canada, in 2013. Since graduation, he has been working in various roles with two major electric utility companies in West Canada as a Professional Engineer (P.Eng.) and also as a Senior Engineer for more than 5 years. His research interests include applications of artificial intelligence and big data technologies in power system planning and operation, power quality data analytics, power equipment testing, and system grounding. Dr. Dong was the recipient of the Certificate of Data Science and Big Data Analytics from Massachusetts Institute of Technology in 2017. He is also a Regional Officer of Alberta Artificial Intelligence Association.

**Alexandre B. Nassif** (S’05–M’09–SM’13) received the doctoral degree from the University of Alberta, Edmonton, AB, Canada. He is currently a Specialist Engineer with ATCO Electric, Edmonton, AB, Canada. Before joining ATCO, he was with Hydro One as a Protection Planning Engineer and with Ryerson University as a Postdoctoral Research Fellow simultaneously. He has authored or co-authored more than 50 technical papers in international journals and conferences in the areas of power quality, DERs, microgrids, and power system protection and stability. Dr. Nassif is a Professional Engineer in Alberta.

**Benzhe Li** (M’18) received the master’s degree from the University of Alberta, Edmonton, AB, Canada, in 2015. He is currently an Electrical Engineer with Energy Ottawa, Ottawa, ON, Canada. His research interests include advanced power quality data analytics and equipment condition monitoring.